\documentclass[12pt]{iopart}
\usepackage{graphics}
\usepackage{epsfig}
\usepackage{graphicx}
\begin{document}

\title{How to Create Black Holes on Earth?}

\author{Marcus Bleicher}

\address{Institut f\"ur Theoretische Physik, J.W. Goethe Universit\"at, 
Max von Laue-Stra\ss{}e 1, 60438 Frankfurt am Main, Germany}
\ead{bleicher@th.physik.uni-frankfurt.de}
\begin{abstract}
We present a short overview  
on the ideas of large extra-dimensions and their implications for the possible production of micro black holes in 
the next generation particle accelerator at CERN (Geneva, Switzerland) from this year on. In fact, the possibility of black hole production on earth is currently
one of the most exciting predictions for the LHC accelerator and would change our current understanding of physics radically.
While it is impossible to discuss the models and implications in full detail here, this article is thought to serve as a 
starting point
for the interested physics students with some basic knowledge about general relativity and
particle physics.
\end{abstract}

\maketitle

\section{Introduction}
Ninety years ago, Karl Schwarzschild, was the first one to present a
solution to Einstein's famous equations for general relativity \cite{Schwarzschild:1916uq}.
His result layed the foundation for the study of some of the most fascinating and mysterious objects known 
today, black holes. Usually one finds these gigantic objects in the centers of galaxies
like our Milky Way. The central black hole in our Milky Way is in the constellation Sagittarius 
(see Ref.\ \cite{saga} for actual pictures of the black hole from the Chandra X-Ray Observatory), has 
a mass of 3 Million Suns and gravitationally binds our galaxy together. Nowadays, it is believed 
that this huge black hole was created from a tiny density fluctuation shortly after the Big Bang \cite{Carr:1975qj}.

Even without detailed knowledge of the theory of general relativity, Schwarzschild's solution can
be easily obtained from Einstein's field equations:
\begin{eqnarray}
R_{\mu \nu} - \frac{1}{2} g_{\mu \nu} R = -  \frac{8\pi}{m_{\rm p}^2}T_{\mu \nu} \quad, \label{EFG}
\end{eqnarray}
where $m_{\rm p}^2 = 1/G$ is the inverse of the Newtonian constant and $m_{\rm p}$ denotes the Planck mass.
The sources of gravity are everything which carries energy and/or momentum. These quantities are described in the
energy momentum tensor $T_{\mu \nu}$ on the right hand side. The sources cause a curvature of space-time which is
described by the metric $g_{\mu \nu}$, and the curvature tensor $R$. For a detailed discussion of the 
Einstein-equations, we refer to Ref. \cite{[book4]}. 
For a spherically symmetric mass distribution the metric outside 
of the mass distribution is then 
described by the Schwarzschild solution \cite{Schwarzschild:1916uq,[book4]}:
\begin{eqnarray} \label{ssmetr}
{\mathrm{d}}s^2= g_{\mu \nu} {\rm d}x^{\mu}  {\rm d}x^{\nu} = -\gamma(r) {\mathrm{d}}t^2 + 
\gamma(r)^{-1}{\mathrm{d}}r^2 + r^2 {\mathrm{d}} \Omega^2 \quad ,
\end{eqnarray}
where 
\begin{eqnarray}  
\gamma(r)=1- \frac{1}{m_{\rm p}^2}\frac{2M}{r} \quad , \label{gamma}
\end{eqnarray}
and ${\mathrm{d}} \Omega$ is the surface element of the 3-dimensional unit sphere and $M$ is the 
total mass of the object. 

Here one observes that black holes are an immediate consequence of the Schwarzschild solution Eq.(\ref{ssmetr}). 
The metric component $g_{tt}=\gamma(r)$ vanishes at the 
radius $R_H = 2 M/m_{\rm p}^2$, the so called Schwarzschild radius. 
However, for usual objects like our sun, this radius is $\approx 2$~Km  and therefore well inside 
the mass distributions where the exterior solution can not be applied. Thus, only if the mass is 
so densely packed that it has a radius smaller than $R_H$, we will have to consider 
the consequences of $g_{tt}\rightarrow 0$. This is the case of black hole formation. 

Astrophysically, the existence of giant black holes is well established - from Schwarzschilds pioneering works to
the calculations of Stephen Hawking, many theoretical and experimental features fit well together.

Just recently, however, new theoretical ideas came up that might completely change our thinking about black holes, 
namely the prediction of ``micro black holes'' \cite{Dimopoulos:2001hw,Bleicher:2001kh}. 
These new approaches for the formation and investigation 
of black holes are based on theoretical models about the microscopic geometry of the universe
broad up over the last couple of years. 

In the following sections I will try to summarize some of the key
ingredients and findings within these new scenarios. The level of presentation is such that a student with 
basic knowledge on general relativity and particle physics should be easily able to follow the line of arguments.
The interested reader is referred to the following review articles for further exploration of (micro) black hole physics 
and large extra dimensions \cite{Cavaglia:2002si,Hossenfelder:2004af,Landsberg:2006mm}.
\section{Extra-dimensions and extremely strong gravity}
The new idea is that our universe might have more than the well known three spatial dimension, height, 
width and length, but up to seven additional space dimensions that are usually unobservable.
These additional dimensions are curled up to tiny ``doughnuts'' to make them hard to observe with a probe
that is bigger than the extension of the ``doughnut'' - i.e. the extra dimensions are compactified on a torus
of radius $R$. 
Surprisingly, in some models these additional dimensions
can be as big a $R \sim 1/100$ millimetre, therefore they are called ``large extra 
dimensions''. These ideas were first put forward by the physicists 
N. Arkani-Hamed, G. Dvali, S. Dimopoulos, I. Antoniadis and L. Randall, R. Sundrum  
in 1998/1999 \cite{Arkani-Hamed:1998rs,Antoniadis:1998ig,Randall:1999ee}. 

What makes these models especially interesting is the fact that the gravitational interaction increases strongly
(by factors on the order of $10^{32}$) on distances below the extension of the extra dimensions.
The strict way to show this is to start from the Einstein-Hilbert action for $d$-dimensional gravity and to
integrate out the additional space like extra dimensions. However, there is an easier way to understand
this. 

Consider a particle of mass $M$ located in a space time with $d+3$ dimensions.
The general solution of Poisson's equation yields its potential as a function 
of the radial distance $r$ from the source
\begin{eqnarray}
\phi(r) = \frac{1}{M_{\rm f}^{d+2}} \frac{M}{r^{1+d}} \quad. \label{newtond}
\end{eqnarray}
Here we have introduced a new fundamental mass-scale $M_{\rm f}$ with a suitable exponent.
Remembering that the additional $d$ space-time dimensions are compactified on radii $R$, 
then, at distances $r \gg R$, the extra dimensions
should be 'hidden' and the potential Eq. (\ref{newtond})
should turn into the well known $1/r$ potential, however with a pre-factor including the volume of the extra 
dimensions
\begin{eqnarray}
\phi(r) \stackrel{r \gg R}{=} \frac{1}{M_{\rm f}^{d+2}} \frac{1}{R^{d}} \frac{M}{r} 
\stackrel{!}{=} \frac{1}{m_{\rm p}^{2}} \frac{M}{r}\quad. \label{newton} 
\end{eqnarray}
In the limit of large distances, this should be identical to  Newton's 
gravitational law, yielding the  relation
\begin{eqnarray}
m_{\rm p}^2 = M_{\rm f}^{d+2} R^d \quad. \label{Master}
\end{eqnarray}
Assuming that $M_{\rm f}$ has the right order of magnitude to be compatible with observed physics,
it can be seen that the volume of the extra dimensions suppresses the 
fundamental scale and thus, explains the huge value of the Planck mass. Today one expects this 
new fundamental scale to be of the order of $M_{\rm f} \sim 1$~TeV, allowing sizes of extra dimensions
between 1/10~mm to 1/1000~fm for $d$ from 2 to 7.
 
\begin{figure}
\includegraphics[width=9cm]{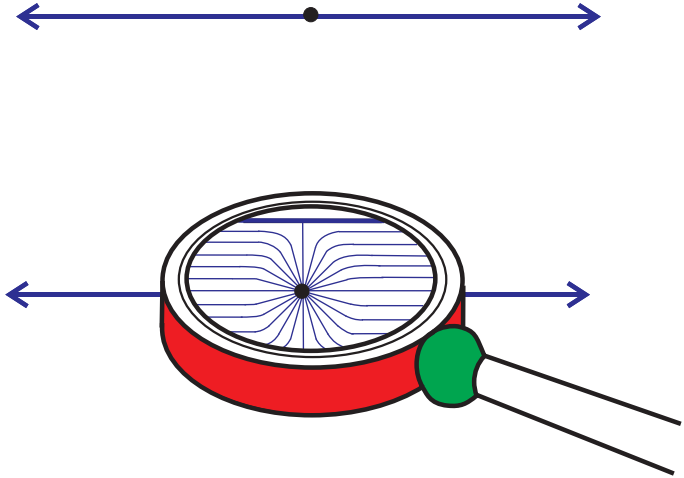}
\caption{\label{fluxlines} Flux lines emerging from a mass point. Top: Along one dimension in usual space time.
Bottom: Spread of the flux lines due to an additional extra dimension. Taken from \cite{Hossenfelder:2004af}.}
\end{figure}

The above arguments can be followed even clearer when one remembers that the strength of interaction is proportional to
the density of flux lines. Fig.\ \ref{fluxlines} illustrates this behaviour: While the top figure shows the
setting with a mass point in the center for one dimension with undiluted flux lines, the lower figure 
shows (in the magnifying glass) that there are actually two dimensions and the field lines are diluted.
Mathematically speaking we find that in the usual three dimensional world the gravitational potential falls-off
like $1/{\rm distance}$, however if we have more dimension (e.g. $d$ additional dimensions) 
to dilute the gravitational flux lines, the potential will fall like $1/{\rm distance}^{d+1}$.

Why is the gravitational field then stronger at small distances? This becomes clear, if we remember that the long
distance behaviour of gravitation is fixed from daily observation (Newton's law). If we extrapolate 
towards smaller distances
the additional dilution of the field lines starts, therefore, to get the proper long distance behaviour back, one
has to start with more dense flux lines (meaning stronger gravitational interaction) at small distances.

Thus, if Newton's law is modified at small distances if extra dimensions are introduced, the most 
obvious experimental test  for the existence of extra dimensions is a 
measurement of the Newtonian potential at small distances. Cavendish like experiments which search
for deviations from the $1/r$ potential have been performed during the last years with high
precision \cite{Newtonslaw} and require the extra dimensions to have radii not larger than $\sim 100\mu$m,
which is the reason to disfavour the case of $d=2$.

Thus, it is interesting to look for phenomena that are sensitive to modifications of the gravitational 
interaction at even smaller distances. With this increase of the gravitational interaction strength, 
also the probability of gravitational 
collapse increases - which is nothing else than black hole formation. In fact, the increase 
in gravitational strength is drastic enough that it might allow black hole production here on earth already 
within the next couple of month. 

\section{Black holes in the laboratory}
How can one imagine the production of black in a particle accelerator? The nuclei of two hydrogen 
atoms (the protons) are
accelerated in opposite directions nearly to the speed of light. When they collide, the subatomic constituents
within the hydrogen nuclei interact. If the distance between these constituents (called quarks and gluons) 
is small enough - i.e. smaller than a thousands of a proton diameter - and the energy is high enough a
micro black hole could be formed \cite{Thorne:1972ji}. The mass of such black holes will be of the 
order of a TeV ($\sim$ five gold atoms)
and therefore be much lighter than the cosmic black holes we know up to now.

This can be strictly derived  by using the higher dimensional Schwarzschild-metric \cite{my}:
\begin{equation}
{\rm d}s^2  = - \gamma (r) {\mathrm d}t^2 + \gamma^{-1}(r) {\mathrm d}r^2 + r^2 {\mathrm d}\Omega^2_{(d+3)}
\quad ,
\end{equation}
where ${\rm d}\Omega_{(d+3)}$ now is the surface element of the 
$3+d$ - dimensional unit sphere and
\begin{equation}
\gamma(r)=1-\left(\frac{R_H}{r}\right)^{d+1} \quad.
\end{equation}
The constant $R_H$ can be obtained by requiring that the solution reproduces the
Newtonian limit for $r\gg R$. This means the derivative of $\gamma$ has to 
yield the Newtonian potential in (3+d) dimensions:
\begin{eqnarray}
\frac{1}{2}\frac{\partial \gamma(r)}{\partial r}=\frac{d+1}{2} \left(\frac{R_H}{r}\right)^{d+1} \frac{1}{r} 
\stackrel{!}{=} \frac{1}{M_{\rm f}^{d+2}} \frac{M}{r^{d+2}} \quad.
\end{eqnarray}
So we have
\begin{eqnarray}
\gamma(r) =  1- \frac{2}{d+1} \frac{1}{M_{\rm f}^{d+2}} \frac{M}{r^{d+1}} \quad,
\end{eqnarray}
and $R_H$ is
\begin{equation} \label{ssradD}
R_H^{d+1}=
\frac{2}{d+1}\left(\frac{1}{M_{\rm f}}\right)^{d+1} \; \frac{M}{M_{\rm f}} \;\; .
\end{equation}
For $M_{\rm f}\sim$1~TeV this radius is $\sim 10^{-4}$~fm. Thus, at high enough energies it
might be possible to bring particles closer together than their horizon, which will force
them to form a black hole. 

\begin{figure}
\epsfig{figure=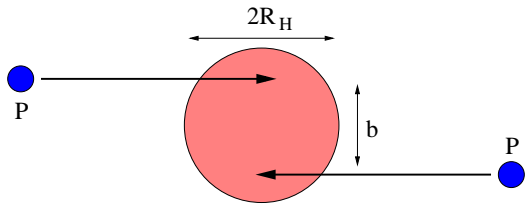,width=7cm }  
\caption{At very high energies, the partons, $p$,
can come closer than the Schwarzschild radius, $R_H$, associated with their energy. If the
impact parameter, $b$, is sufficiently small, such a collision will inevitably generate
a black hole.}
\label{bumm}
\end{figure}

Let us now calculate the possible production rate of black holes explicitly. I.e., we 
consider two elementary particles, approaching each other with a very high kinetic energy in
the c.o.m. system slightly above the new fundamental scale $M_{\rm f}\sim 1$~TeV, 
as depicted in Figure \ref{bumm}. Since the black hole 
is not an ordinary particle of the Standard Model
and its correct quantum theoretical treatment is unknown, it is treated as a meta-stable
state, which is produced and decays according to the semi classical formalism of black
hole physics.

To compute the formation probability, we approximate the cross section of the black holes 
by their classical geometric area 
\begin{eqnarray} \label{cross}
\sigma(M)\approx \pi R_H^2 \Theta(M-M_{\rm f})\quad.
\end{eqnarray}
with $\Theta$ being the Heaviside function\footnote{The Heaviside function is defined as $$\Theta(x) = 
\left\{ 
\begin{array}{lr}
0&x\le 0\\
1&x> 0
\end{array} 
\right. $$ and assures that black hole production is only possible above the Planck 
mass or the new fundamental mass $M_{\rm f}$, resp.}.
Setting $M_{\rm f}\sim 1$~TeV and $d=2$ one finds 
$\sigma \sim 400$~pb.
Using the geometrical cross section formula, it is now possible to compute the differential cross section
${\mathrm d}\sigma/{\mathrm d}M$ which will tell us how many black holes will be formed with a certain
mass $M$ at a c.o.m. energy $\sqrt{s}$. 
The probability that two colliding particles will form a black 
hole of mass
$M$ in a proton-proton collision at the {\sc LHC} (at a center of mass energy of 14~TeV) involves 
the parton distribution functions. These
functions, $f_A(x,\hat{s})$, parametrise the probability of finding a constituent $A$ of the
proton (quark or gluon) with a momentum fraction $x$ of the total energy of the proton. 
These constituents are also called partons. Here,
$\sqrt{\hat{s}}$ is the c.o.m. energy of the parton-parton collision. For details about the
calculations of scattering cross sections from parton distribution functions, the reader is referred to \cite{hm}.

The differential cross section is then given by summation over all possible parton interactions and
integration over the momentum fractions $0\le x_1\le 1$ and $0\le x_2\le 1$, where the 
kinematic relation $x_1 x_2 s=\hat{s}=M^2$ has
to be fulfilled. This yields
\begin{eqnarray} \label{partcross}
\frac{{\rm d}\sigma}{{\rm d}M}  
&=&  \sum_{A_1, B_2} \int_{0}^{1} {\rm d} x_1 \frac{2 \sqrt{\hat{s}}}{x_1s} f_A(x_1,\hat{s}) 
f_B (x_2,\hat{s})  \sigma(M,d)   \quad.
\end{eqnarray}
The particle distribution functions for $f_A$ and $f_B$ are tabulated e.g. in the CTEQ-tables \cite{cteq}.
One might think that the mass distribution diverges when $x_1=0$, however, this limit does not fulfill the kinematic
constrained $x_1x_2=M^2/s$, because it would result in $x_2>1$ for finite black hole masses $M\ge M_{f}$.

\begin{figure}
\epsfig{figure=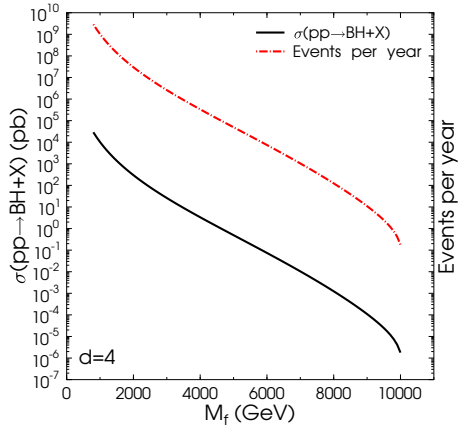,width=7cm }  
\caption{Black hole yield per year and production cross section in pp interactions at a c.o.m. energy of
14~TeV.}
\label{sigma}
\end{figure}

It is now straightforward to compute the total cross section and number by integration over Eq. (\ref{partcross})
This also allows us to estimate the total number of black holes, $N_{\rm BH}$, 
that would be created at the {\sc LHC} per year. It is $N_{\rm BH}/{\rm year}= \sigma(pp \to {\rm BH}) L$
where we insert the estimated luminosity for the {\sc LHC},  $L=10^{33}{\rm cm}^{-2}{\rm s}^{-1}$. This
yields at a c.o.m. energy of $\sqrt{s}=14$~TeV the total number of 1 billion black holes per year! 
This means, about ten black holes per second would be created. The total number of black holes produced per year
at the LHC as a function of the new fundamental mass (being related to the size of the extra dimensions) is depicted
in Fig. \ref{sigma}.
The importance of this process led to a high number of publications on the topic of TeV-mass black holes at
colliders both in scientific journals \cite{Dimopoulos:2001hw,Bleicher:2001kh,Giddings3,Cavaglia:2004jw,Atlas,BHex,BHlep} and in the popular press \cite{nyt}. 

Especially for teachers and young students it should be noted that numerical simulation and visualisation 
packages for the production and decay of black holes in colliders and from cosmic rays are available. 
These packages are currently used to explore potential signatures for black hole production and go by the
names CHARYBDIS \cite{Harris:2003db}, GROKE \cite{groke,Ahn:2005bi} and CATFISH \cite{Cavaglia:2006uk}. 

\section{Discussion}

The final question that remains is: Are these micro black holes dangerous?
Up to now it is still an open question whether these tiny black holes are stable or decay. 
The British physicist Stephen Hawking predicted thirty years ago that black holes should decay in a 
quantum mechanical process known as the Hawking radiation \cite{Hawk1}. The temperature of this radiation from black holes
can be calculated following Hawking's arguments and is given by:
\begin{eqnarray} \label{kappa}
T = \frac{\kappa}{2 \pi} \quad, 
\end{eqnarray}
where $\kappa= 1/2\, \partial\gamma(r)/\partial r$ at $r=R_H$ is the surface gravity of the black hole.
For the case of black holes in extra dimensions one obtains $\kappa= (1+d)/2\cdot 1/R_H$.
Putting numbers in, yields temperatures on the order of several hundred GeV. I.e. the
black hole will explode immediately into all available (standard model) particles. The most prominent
signatures of this explosion would be an event with multiple quark- and gluon-jet production without
back-to-back correlation (as would be expected from processes of the strong interaction).

However, it is unclear whether these 
calculations are also valid for such micro black holes. In fact, it is currently discussed that micro black 
holes can form stable remnants that might constitute some new form of stable dark matter.
If these micro black holes do radiate, they can be seen in the detectors at CERN due to their extremely
high (Hawking-) temperature of many billion Fahrenheit.

If the black holes are stable, they will fall towards the center of the earth and will be hidden
in the earth's core. 
The probability that a black hole remnant starts to grow inside the earth is proportional to 
its volume and therefore extremely small. It would take many of years before 
another particle will be caught by this remnant.
In addition one can imagine that ultra high energetic cosmic rays (UHECR) might have produced micro black holes 
either by collisions with the earth's material or the atmosphere. Assuming a reasonable flux of cosmic neutrinos 
and their interaction with earth material over the whole existence
of the earth leads to the surprising result that the earth might have already been hit by nearly a pound
of black holes. The production of mini black holes from UHECRs has led to 
many exciting predictions over the recent years \cite{Feng:2001ib,Anchordoqui:2001ei,Anchordoqui:2001cg,Ringwald:2001vk,Ahn:2005bi}. Basically, these studies have found black hole production rates due to cosmic ray interactions
with the earth and its atmosphere ranging from one black hole per year to one black hole per day. 
Since cosmic rays have been hitting the earth for billions of years  (possibly resulting in the production of  
billions of micro black holes) it can be seen as strong experimental evidence that the predicted micro black holes 
do not pose a threat.
 
While nothing is ever known completely before one has experimentally explored it, there 
is no scenario known up to now that leads to any dangerous implications 
for mankind from the production of micro black holes on earth.

\subsection*{Fiction or reality?}

Let me finally discuss the likelihood for the production and observation of black holes in future 
experiments. It should be clear that several (huge) assumptions have to be fulfilled to make the
above predictions a reality:
\begin{itemize}
\item
Space time has to have additional space like dimensions to allow for a decrease of the Planck scale to
an experimentally accessible value.  While String Theory indeed predicts a multi-dimensional 
universe, based on theoretical grounds about anomaly cancellations, the existence of additional 
dimensions is still a speculation today without experimental proof.

\item
The size of the extra dimensions has to be ''large''. Meaning that if the dimensions extensions are below 
$10^{-4}-10^{-5}$~fm they can not be probed by near future accelerators. The spatial resolution probed by
interactions is at best $1/\sqrt s$, and therefore $\sim 10^{-4}$~fm in proton-proton collisions at LHC
and $\sim 10^{-6}$~fm for the highest energetic cosmic rays. Unfortunately, many models (not the ADD-model 
discussed here) assume only ''small'' extra dimensions with sizes around the usual Planck length.

\item
The unknown new fundamental scale has to be below some ten TeV to allow for collider studies. 

\item
Even if micro black holes are formed, it might be difficult to detect them, because the emission 
pattern of small black holes near the Planck mass might be strongly altered from our expectations.
\end{itemize}

\section{Summary}
The aim of this article was to give a brief introduction on recent new ideas about the
geometry of space. A spectacular implication of these new models with additional dimensions is
the increased production probability of black holes. While it might sound like science fiction
the next generation particle accelerators will be able to test this prediction.
The experimental program at LHC will start end of the year and might for the first time allow to explore 
the properties of black holes here on earth.

\end{document}